\documentclass[prr,aps,twocolumn,10pt,superscriptaddress,notitlepage,longbibliography]{revtex4-1}
\usepackage[margin=1.54cm]{geometry}
\usepackage[caption=false]{subfig}
\usepackage{graphicx}
\usepackage{amsmath,mathtools,bm}
\usepackage{amssymb}
\usepackage{epstopdf}
\usepackage{siunitx}
\usepackage{color}
\usepackage[ngerman,english]{babel}

\usepackage{lipsum, babel}

\usepackage{tabularx}

 \definecolor{winered}{rgb}{0.5,0,0}
 
\usepackage[colorlinks]{hyperref}
 \hypersetup{
 colorlinks=true,
 linkcolor=winered,
 urlcolor={winered},
 filecolor={winered},
 citecolor={winered},
 allcolors={winered}
 }

\usepackage{braket}

\usepackage{letltxmacro}
\LetLtxMacro{\ORIGselectlanguage}{\selectlanguage}
\makeatletter
\DeclareRobustCommand{\selectlanguage}[1]{%
  \@ifundefined{alias@\string#1}
    {\ORIGselectlanguage{#1}}
    {\begingroup\edef\x{\endgroup
       \noexpand\ORIGselectlanguage{\@nameuse{alias@#1}}}\x}%
}
\newcommand{\definelanguagealias}[2]{%
  \@namedef{alias@#1}{#2}%
}

\DeclareSymbolFont{matha}{OML}{txmi}{m}{it}
\DeclareMathSymbol{\varv}{\mathord}{matha}{118}

\usepackage{amsmath}
\newcommand\ddfrac[2]{\frac{\displaystyle #1}{\displaystyle #2}}
\usepackage{graphicx,wrapfig}

\usepackage{amsmath}
\usepackage{float}
\makeatother

\definelanguagealias{en}{english}
\definelanguagealias{EN}{english}
\definelanguagealias{eng}{english}
\definelanguagealias{de}{ngerman}
\usepackage[utf8]{inputenc}

\begin{document}


\title{Quasinormal mode theory of elastic Purcell factors
and Fano resonances of optomechanical beams}

\date{\today}
\author{Al-Waleed El-Sayed}
\email{14awes@queensu.ca}
\affiliation{Department of Physics, Engineering Physics and Astronomy,
Queen's University, Kingston, ON K7L 3N6, Canada}
\author{Stephen Hughes}
\email{shughes@queensu.ca}
\affiliation{Department of Physics, Engineering Physics and Astronomy,
Queen's University, Kingston, ON K7L 3N6, Canada}

\begin{abstract}
   We introduce a quasinormal mode theory of mechanical open-cavity modes for optomechanical resonators, and demonstrate the importance of using a generalized (complex) effective mode volume and  the phase of the quasinormal mode. We first generalize 
   and fix 
   the normal mode theories of the elastic Purcell factor, and then show a striking example of coupled quasinormal modes yielding a pronounced Fano resonance. Our theory is exemplified and confirmed by full three-dimensional calculations on optomechanical beams, but the general findings apply to a wide range of mechanical cavity modes. This quasinormal mechanical mode formalism, when also coupled with a quasinormal theory of optical cavities, offers a unified framework for describing a wide range of optomechanical structures where dissipation is an inherent part of the resonator modes.
\end{abstract}

\maketitle

\section{Introduction}
\label{sec:introduction}

The ability to describe optical cavities in terms of normalized modes and cavity figures of merit
has played a significant role in laser optics and cavity quantum electrodynamics (cavity-QED). Mode theories
not only quantify the underlying physics, but they simplify the numerical modelling 
requirements and are essential for defining quantized modes in quantum field theory. A striking example
is the Purcell factor \cite{purcell_resonance_1946},
which elegantly describes the enhanced emission rate
of a quantum dipole emitter:
\begin{equation}
F_{\rm P} = \frac{3}{4\pi^2}\left(\frac{\lambda_0}{n}\right)^2 \, \frac{Q}{V_{\rm eff}},    
\end{equation}
where $\lambda_0$ is the free space wavelength,
$n$ is the refractive index,
$Q$ is the quality factor, and ${V}_{\rm eff}$
is the effective mode
volume. Purcell's  theory was originally derived
for closed cavity systems, though loss is partly accounted for in the definition
of $Q$. The above formula assumes
perfect spatial and polarization alignment of the emitter, which is typically achieved at a field antinode.

Recently, a corrected form for Purcell's formula has been
derived in terms of quasinormal modes (QNMs)~\cite{kristensen_generalized_2012,kristensen_modeling_2020}, which are the modes of an
open cavity resonators with complex eigenfrequencies; the QNMs yield
a generalized (or complex) mode volume~\cite{kristensen_generalized_2012,kristensen_modes_2014}, $\tilde V_{\rm eff}$, and only the 
real part is used in Purcell's formula. This subtle ``fix'' can have profound
consequences, and applies to a wide range of lossy cavity structures, including plasmonic resonators~\cite{Sauvan2013,Ge2014njp}
and hybrid structures of metals and dielectric parts~\cite{kamandar_dezfouli_modal_2017}. Moreover, very recently, it was also 
shown how to quantize these QNMs~\cite{franke_quantization_2019,PhysRevResearch.2.033456}, where the dissipation becomes an essential component in explaining the breakdown of the Jaynes-Cumming model for several modes,  causing intrinsic quantum mechanical coupling between classically orthogonal modes.

There are significant analogies
between optics and acoustics/mechanics, where the wave equations for acoustics is given in terms
of the pressure and velocity fields~\cite{PhysRevB.99.174310},
instead of the electromagnetic fields for optics. In typical resonator structures,
both systems yield open cavity modes with complex eigenfrequencies.
Moreover, optomechanical structures can support coupling between
mechanical and optical modes~\cite{eichenfield_picogram-_2009}, offering a wide range of applications in optomechanics~\cite{aspelmeyer_cavity_2014,aspelmeyer_quantum_2010,kippenberg_cavity_2007,favero_optomechanics_2009,verhagen_quantum-coherent_2012,Weis2010}.
Despite these similarities, the common use of cavity mode theories
in optics is less developed in elastics, and, for example, one rarely talks about ``mechanical''
effective mode volumes.

In optomechanical systems, the radiation forces exerted by photons are exploited to induce, control, and/or measure mechanical motion in resonators over a wide range of length scales. The applications of optomechanics vary widely~\cite{aspelmeyer_cavity_2014}, from the transduction~\cite{balram_coherent_2016} and storage~\cite{fiore_storing_2011,bagheri_dynamic_2011} of quantum information, to ultra-sensitive mass sensing~\cite{yu_cavity_2016,liu_sub-pg_2013}. 
Other applications include ground-state cooling~\cite{Lau2020} and
nonlinear optomechanics~\cite{Sankey2010}.
The stereotypical optomechanical system consists of a laser-driven cavity whose electromagnetic fields exert a radiation pressure force on a mechanical resonator, which then acts back on the cavity mode, causing the two modes to interact. Modern optomechanical systems can take the form of ultra-thin membranes, micro-ring resonators, and nano-structures acting a  photonic crystal~\cite{yablonovitch_inhibited_1987,krauss_two-dimensional_1996} and a phononic crystal~\cite{kushwaha_acoustic_1993,sigalas_band_1993,maldovan_sound_2013}  simultaneously~\cite{laude_phoxonic_2016,rolland_acousto-optic_2012,el_hassouani_dual_2010,djafari-rouhani_phoxonic_2016,kipfstuhl_modeling_2014}, which have been shown to have direct applications in on-chip quantum information processing~\cite{eichenfield_modeling_2009,chan_optimized_2012,kalaee_design_2016,pitanti_strong_2015,krause_high-resolution_2012,winger_chip-scale_2011}. {Superconducting circuits have also recently been shown to exhibit optomechanical-like coupling~\cite{johansson_optomechanical-like_2014}}.

\begin{figure}[thb]
    \centering
    \includegraphics[width=0.48\textwidth]{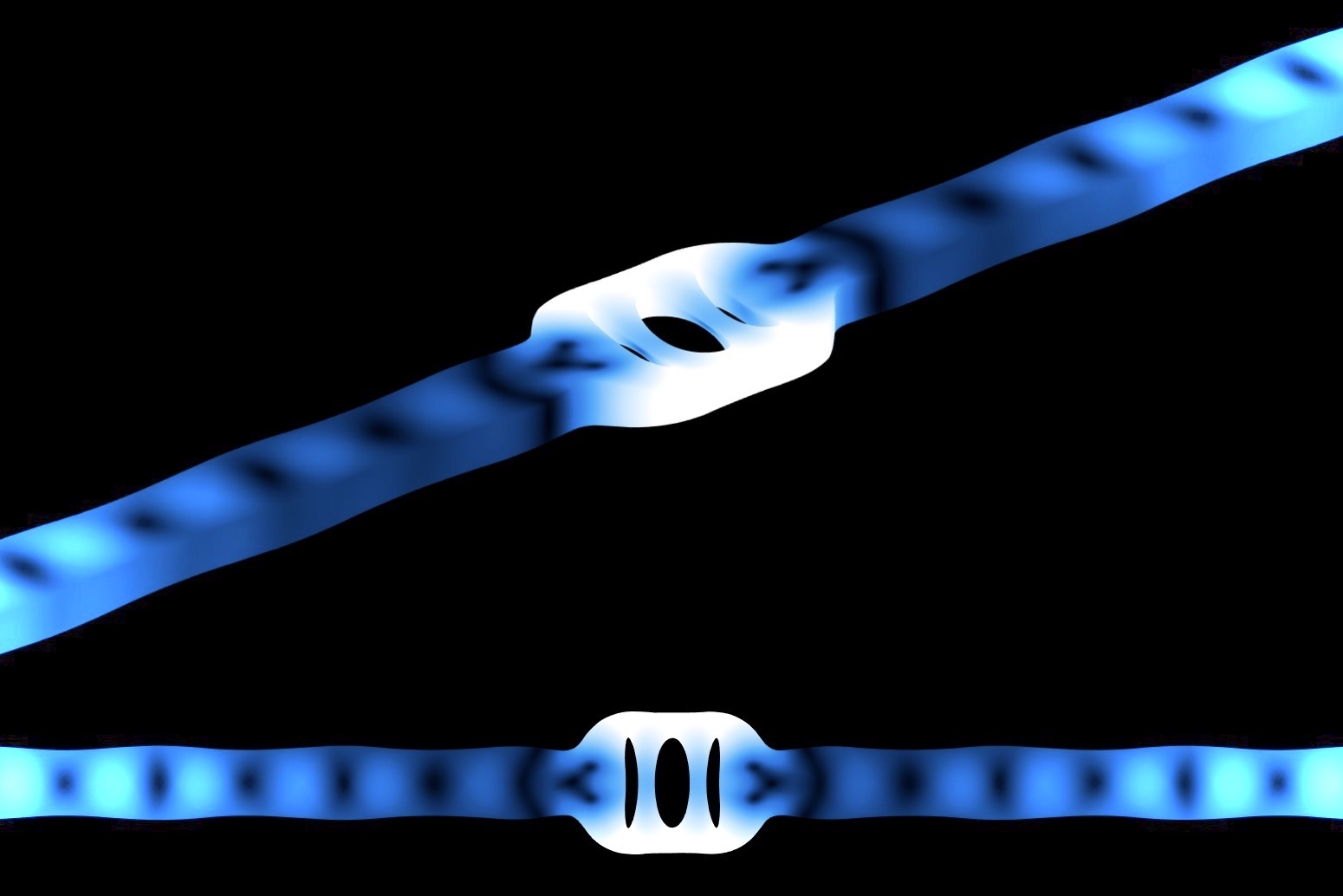}
    \caption{Visualization of a mechanical QNM on a 3-hole optomechanical nanobeam cavity, which will spatially diverge because of teh outgoing boundary conditions.}
    \label{fig:theory}
\end{figure}

In most optomechanical mode theories to date,
the optomechanical coupling rate $g_{0}$ is rarely taken from a first-principles model, which is in contrast
to {\em modal} methods in optics where it is more common to adopt an analytical approach based on the optical modes of the structure. Yet there is clearly a need to describe 
emerging effects such as mode-to-mode transcription
and reservoir engineering
in terms of  
the underlying mode properties of the
mechanical cavity modes, both in   
classical and quantum mechanical problems.
A recent theory paper introduced the interesting idea of an {\it elastic} Purcell factor~\cite{schmidt_elastic_2018}, which like its optical counterpart, describes an enhancement of a dipole emitter (but now a {\em force dipole}), in terms of  $Q$ and $V_{\rm eff}$; similar to earlier works on optical cavities, the authors used a ``normal mode theory,'' which is in general incorrect for open cavity modes~\cite{kristensen_generalized_2012};
however, for high $Q$ cavities, the
normal mode approach can be a very good
approximation, but the theory is still ambiguous.
Experimental measurements have also recently been performed on the
acoustic Purcell factor~\cite{landi_acoustic_2018}, and thus there is clearly a need to develop mode theories for such geometries and emerging material systems beyond optics.


To account for mode dissipation in
optical resonators, the modelling of the dominant cavity mode is usually calculated by implementing outgoing boundary conditions (otherwise it has an infinite lifetime).
Often this problem is treated with closed boundary conditions
or as a Hermitian eigenvalue problem, but this is  inconsistent with
a finite cavity loss.
In fact, it is now known that all open cavity modes yield spatially divergent modes, which are the QNMs described earlier. Figure \ref{fig:theory} shows a schematic representation of an elastic QNM 
from an optomechanical photonic crystal beam (calculated in detail later), and we note that the mode diverges 
for spatial position far down the open beam.  These QNMs
 have recently proven to be very powerful in photonics design and simulations~\cite{kristensen_modes_2014,lasson_semianalytical_2015,lalanne_light_2018,kristensen_modeling_2020}. For the purpose of field normalization and developing mode theories, both optical and mechanical modes are usually assumed to be lossless and then dissipation is added later through system-bath coupling theory, or phenomenologically; in contrast, the QNM approach includes losses from the beginning since the eigenfrequencies are complex, unlike normal modes. The QNMs also quantify coupling parameters in a more complete way, an example of which has been shown for 
 two coupled QNMs of dielectric-cavity systems~\cite{kamandar_dezfouli_modal_2017}, resulting in striking interference effects that demonstrate how the phase of the mode must be maintained.

In this work, we introduce an intuitive and accurate
QNM description for mechanical modes
$\tilde {\bf q}_{\rm m}$, which have complex
eigenfrequencies,
$\tilde{\Omega}_{\rm m}=\Omega_{\rm m}-i
\gamma_{\rm m}$, and
$Q_{\rm m}=\Omega_{\rm m}/2\Gamma_{\rm m}$. For single mode resonators,
This QNM formalism
allows a rigorous definition of the effective mode volume $V_{\rm eff,m}$~\cite{purcell_resonance_1946,kristensen_generalized_2012,kristensen_modes_2014} for mechanical modes~\cite{eichenfield_modeling_2009}, or,
equivalently in mechanics, the effective motional mass $m_{\rm eff,m} = \rho V_{\rm eff,m}$~\cite{pinard_effective_1999,gillespie_thermally_1995,eichenfield_modeling_2009}, which is more commonly used in the optomechanics literature~\cite{kipfstuhl_modeling_2014,li_optomechanical_2015,zheng_femtogram_2012,chan_optimized_2012,kalaee_design_2016,djafari-rouhani_phoxonic_2016}.
We first  present a {\it generalized} elastic effective mode volume $\tilde{V}_{\rm eff,m}(\mathbf{r})$ using a QNM normalization, and show 
the problems with using a normal mode volume 
 $V_{\rm eff}$. We then use this complex, position dependant $\tilde{V}_{\rm eff,m}(\mathbf{r})$  to carry out an analytical Green function expansion~\cite{Ge2014njp}, which can be used to 
 quickly solve a wide range of force-displacement problems in an analogous way to how the photon Green function is being used to carry out light-matter investigations in optics~\cite{yao_-chip_2010}.
 Then, 
using the case of two coupled modes,
 we demonstrate the accuracy of the QNM theory in explaining complex Fano resonances,
 and demonstrate the clear failure of the usual normal mode theory for these acoustic modes.



{A Fano resonance is a well-known scattering phenomenon that results in asymmetric spectral lineshapes. They have been shown to exist in various physical systems, finding applications in photonics \cite{limonov_fano_2017}, plasmonic metamaterials \cite{khanikaev_fano-resonant_2013,lukyanchuk_fano_2010}, and optomechanics \cite{qu_fano_2013,abbas_investigation_2019}. In optics, systems that exhibit these phase-dependant interference effects may provide new methods of manipulating light propagation.
 Applications include sensing, lasing, and optical switching \cite{Heuck2013,lukyanchuk_fano_2013,chen_fano_2015,limonov_fano_2017}. Experimental evidence of Fano resonances in coupled nanomechanical resonators has also  been demonstrated in \cite{stassi_experimental_2017}. Fano resonance phenomena in optomechanical systems can potentially be used for processing classical and quantum information, where the hybridized mechanical modes exhibiting Fano excitation lineshapes allow for an on-chip platform for storage and photonic-phononic quantum state transfer,
 demonstrated experimentally by studying the coherent mixing of mechanical excitations within optomechanical cavities~\cite{lin_coherent_2010}. 
 Often Fano-resonances are associated with the interference between a bound resonance and a continuum, such as through a  cavity and waveguide, but two coupled resonators can also
 yield a Fano resonance.
 For example,
 hybrid plasmonic-dielectric systems cam show a significant Fano resonance~\cite{Barth2010,Doeleman2016,kamandar_dezfouli_modal_2017},
which can be explained using optical QNM theory~\cite{kamandar_dezfouli_modal_2017} with only two coupled QNMs,  and also gives rise to new regimes in quantum optics~\cite{franke_quantization_2019, PhysRevResearch.2.033456}.
 The investigation of coupled mechanical modes in optomechanical systems often rely on the power spectral density measurements with finite element methods typically being used for only initial rudimentary descriptions,
 so there is clearly motivation in being able to describe such effects  as the level of an intuitive and accurate mechanical mode theory. }

The rest of our paper is organized as follows:
In Sec.~\ref{sec:theory}, we present the main theory details and important formulas, introducing the elastic
wave equations, QNMs, generalized effective mode volume, and
elastic Purcell formula.
In Sec.~\ref{sec:results}, we present
numerical calculations for a
fully 3D optomechanical beam,
first for a single QNM design, and then for coupled QNMs
that show a striking Fano resonance. In 
both cases, we highlight the 
failure of using a normal mode theory, which is shown to be drastic in the case of two coupled modes. 
Finally, in Sec.~\ref{sec:conclusions}, we present
closing discussions and our main conclusions.

\section{Theory}\label{sec:theory}

\subsection{Wave equations, normal modes, quasinormal modes and Green functions}

Vibrational modes of solids can be calculated using 
the linear theory of elasticity \cite{Slaughter2002}, where one 
assumes infinitesimal deformations and stress forces that do not result in ``yielding'' (deformation point of no elastic return). 

We first
express Newton's second law in terms of the displacement vector
$\mathbf{u}({\bf r},t)$:
\begin{equation}
\label{eq:motion3}
{\bm \nabla} \cdot {\bm \sigma}(\mathbf{r},t) - \rho(\mathbf{r})\ddot{\mathbf{u}}(\mathbf{r},t) = - \mathbf{f}(\mathbf{r},t),
\end{equation}
where $\rho$ is the mass per unit volume,
$\mathbf{f}(\mathbf{r},t)$ is the force
vector or force per unit volume,
and ${\bm \sigma}$ is the stress tensor.
Assuming a harmonic solution of the form,
${\bf u}({\bf r},t)=\mathbf{u}({\bf r},\Omega)e^{-i\Omega t}$, 
and in the absence of a force excitation,
 we obtain the wave equation
in an analogous form to the Helmholtz equation
in optics:
\begin{equation}
\label{eq:n2}
{\bm \nabla} \cdot{\bm \sigma}(\mathbf{r},\Omega) + \Omega^2\rho(\mathbf{r})\mathbf{{u}}(\mathbf{r},\Omega) = 0.
\end{equation}
The stress tensor can be expressed 
as \cite{snieder_chapter_2002}
\begin{equation}
\label{eq:stress}
{\sigma_{ij}(\mathbf{r},\Omega)} = c_{ijkl}\partial_{k}u_{l}(\mathbf{r},\Omega),
\end{equation}
where $c_{ijkl}$ is the fourth-order elasticity tensor (where one sums over repeated indices as with Einstein summation convention), and  $\partial_k \equiv \frac{\partial}{\partial x_k}$ is the partial derivative with respect to the $x_k$ direction. 
Subjecting Eq.~\eqref{eq:n2} to periodic or hard-wall boundary conditions would 
yield ``normal modes'' $\mathbf{q}_{{\rm m}}$ from the following eigenvalue problem:
\begin{equation}
\label{eq:wq}
\rho(\mathbf{r}){\Omega}_{\rm m}^2{q}_{{\rm m}\it{i}}(\mathbf{r}) {+} {\partial}_{j}(c_{ijkl}\partial_{k}{q}_{{\rm m}\it{l}}(\mathbf{r})) = 0,
\end{equation}
where the eigenfrequencies are real and do not account for dissipation.

It is also useful to 
define the corresponding Green function ${\mathbf{G}}_{\rm}$, obtained from~\cite{snieder_chapter_2002}:
\begin{multline}
\label{eq:eomG}
 \partial_{j}(c_{ijkl}\partial_{k}G_{ln}(\mathbf{r},\mathbf{r'};\Omega)) \\  + \rho(\mathbf{r})\Omega^2G_{in}(\mathbf{r},\mathbf{r'};\Omega) = -\delta_{in}\delta(\mathbf{r} - \mathbf{r'}),
\end{multline}
where the $i$th component
of a unit force in the
$n$ direction
at location ${\bf r}'$
is given by $\delta_{in}
\delta({\bm r}-{\bm r}')$.
The displacement can be written in terms of ${\mathbf{G}}_{\rm}$ as \cite{snieder_chapter_2002}:
\begin{multline}
\label{eq:uGfull}
{ u_i}(\mathbf{r}) = \int_{V} {G_{in}}(\mathbf{r},\mathbf{r'})\cdot{f_n}(\mathbf{r})d{\bf r'} \\
+  \oint_S \{ G_{in}(\mathbf{r},\mathbf{r'}) \hat {\rm s}_j c_{njkl}\partial_{k}'u_{l}(\mathbf{r'}) \\ - u_{n}(\mathbf{r'})
\hat {\rm s}_j c_{njkl}\partial_{k}'G_{il}(\mathbf{r,r'})\}d{\bf r'} , 
\end{multline}
where $\hat{\rm s}_j$  is the normal vector of $S$, {the surface forming the outer boundary of the elastic body}. For an elastic body surrounded by empty space, the second term vanishes as there is no stress at the surface. In the case of a point force excitation, $\mathbf{f}(\mathbf{r},t) = \mathbf{f}_{\rm d}\delta({\bm r}-{\bm r}_0)
\delta(t)$, then 
\begin{equation}
\label{eq:uG}
{\mathbf{ u}}(\mathbf{r},\Omega) = \mathbf{G}(\mathbf{r},\mathbf{r}_0; \Omega)\cdot{\bf f}_{\rm d}(\mathbf{r}_0) ,
\end{equation}
where ${\bf f}_{\rm d}$ is a point force at position $\mathbf{r}_0$.

Using the completeness relation for normal modes,
\begin{equation}
\label{eq:complet}
{\mathbf{I}}\delta(\mathbf{r} - \mathbf{r'}) = \sum_{{\rm m}=1,2,\cdots}\rho(\mathbf{r}){\mathbf{ q}}_{\rm m}(\mathbf{r}){\mathbf{ q}}^{\dagger}_{\rm m}(\mathbf{r'}),
\end{equation}
(where $\mathbf{I}$ is a unit dyadic of a 3x3 matrix, {the dagger denotes
the complex conjugate of the transpose: $\mathbf{q}^\dagger = (\mathbf{q}^T)^*$}, and ${\rm m}=1,2, \cdots$)
the Green function can be obtained from an expansion over the
normal modes~\cite{snieder_chapter_2002}
\begin{equation}
\label{eq:GF1}
{\mathbf{G}}(\mathbf{r},\mathbf{r'};\Omega)  = \sum_{{\rm m}=1,2,\cdots}
\frac{\mathbf{q}_{\rm m}(\mathbf{r}) \mathbf{q}_{\rm m}^\dagger(\mathbf{r'})} {\Omega^2_{\rm m} - \Omega^2}.
\end{equation}

For problems in cavity physics with a few modes of interest, the above
theory is not that useful or practical, since one needs
a continuum of modes. Instead, similar to 
open-cavity problems
in optics, one desires to describe the main physics
in terms of just a few discrete resonator modes.
In an elastic resonator medium with open boundary conditions,
the resonator modes $\tilde{\mathbf{q}}_{\rm m}$ are obtained from:
\begin{equation}
\label{eq:wq}
\rho(\mathbf{r})\tilde{\Omega}_{\rm m}^2\tilde{q}_{{\rm m}\it{i}}(\mathbf{r}) { + }  {\partial}_{j}(c_{ijkl}\partial_{k}\tilde{q}_{{\rm m}\it{l}}(\mathbf{r})) = 0,
\end{equation}
where $\tilde{\Omega}_\mathbf{{\rm m}}$ is the complex eigenfrequency
and $\tilde {\bf q}_{\rm m}({\bf r})$ are 
the QNMs.

We can now exploit techniques that
have been recently developed
for obtaining QNMs and QNM Green functions
in  optics~\cite{kristensen_modes_2014,lee_dyadic_1999,lalanne_light_2018,kristensen_modeling_2020}. 
Using a modified completeness relation for QNMs
\cite{lee_dyadic_1999}
\begin{equation}
\label{eq:complet}
{\mathbf{I}}\delta(\mathbf{r} - \mathbf{r'}) = \sum_{{\rm m}=\pm1,\pm2,\cdots}\frac{1}{2}\rho(\mathbf{r})\tilde{\mathbf{ q}}_{\rm m}(\mathbf{r}){[}\tilde{\mathbf{ q}}_{\rm m}(\mathbf{r'}){]^{T}},
\end{equation}
where now ${\rm m}=\pm1,\pm2, \cdots$, superscript $T$ denotes the transpose, and we 
assume this condition is 
 satisfied for spatial positions within or near the cavity~\cite{kristensen_modes_2014,lalanne_light_2018}. {Note that the tensor product in Eq.~\ref{eq:complet} is also frequently written as $\tilde{\mathbf{ q}}_{\rm m}(\mathbf{r})\otimes\tilde{\mathbf{ q}}_{\rm m}(\mathbf{r'})$}.
 We can thus
 expand ${\mathbf{G}}$ in terms of QNMs, through {(alternative expansion forms, for optical QNMs, are discussed in \cite{kristensen_modeling_2020})}
\begin{equation}
\label{eq:greenAns}
{\mathbf{G}}(\mathbf{r},\mathbf{r'};\Omega) = \sum_{{\rm m}=\pm1,\pm2,\cdots}\frac{\tilde{\mathbf{q}}_{\rm m}(\mathbf{r}){[}\tilde{\mathbf{q}}_{\rm m}(\mathbf{r'}){]^{T}}}{2{\Omega}(\tilde{\Omega}_{\rm m} - {\Omega})}.
\end{equation}

To have quantities that better relate
to a mode volume in optics, which is also a key quantity
in Purcell's formula, we next define
an alternative
QNM through 
\begin{equation}
\label{eq:U}
\tilde{\mathbf{Q}}_{{\rm m}}(\mathbf{r}) = \sqrt{\rho({\bf r})}\,\tilde{\mathbf{ q}}_{{\rm m}}(\mathbf{r}), 
\end{equation}
so that $\tilde{\mathbf{ Q}}^2_{{\rm m}} $ has units of inverse volume. 
There have been various approaches
to obtain normalized QNMs in optics~\cite{kristensen_generalized_2012,kristensen_modes_2014,kristensen_normalization_2015,Sauvan2013,lalanne_light_2018,muljarovPert,bai_efficient_2013,PhysRevA.98.043806},
and in this work we use 
\begin{multline}
\label{norm}
  \langle\langle{\tilde{\mathbf{Q}}}_{{\rm m}}|{\tilde{\mathbf{Q}}_{{\rm n}}}\rangle\rangle = \lim_{V\to\infty} \int_{V}  \tilde{\mathbf{Q}}_{{\rm m}}(\mathbf{r})\cdot\tilde{\mathbf{Q}}_{{\rm n}}(\mathbf{r})d \mathbf{r}  \\ + i\ddfrac{{\varv}_{\rm s}^{\rm b}}{4\pi\tilde\Omega_{\rm m}} \int_{{A}} \tilde{\mathbf{Q}}_{{\rm m}}(\mathbf{r})\cdot\tilde{\mathbf{Q}}_{{\rm n}}(\mathbf{r}) d \mathbf{r} = \delta_{{\rm m},{\rm n}},
\end{multline}
which is analogous to the optical QNM normalization introduced by Lai {\it et al.} \cite{lai_time-independent_1990}, 
and was used to introduce the idea of a
``generalized mode volume'' with optical cavities~\cite{kristensen_generalized_2012}.
Compared to its optical counterpart, we note the
 following substitutions: (i)
    speed of light, $c$ $\rightarrow$ shear  speed of sound in bulk material ${\varv}_{\rm s}^{\rm b}$ (as we are only considering transverse modes); and (ii), refractive index $n_{\rm B}$ $\rightarrow$ material density $\rho$.
Note that Eq.~\eqref{norm} needs a careful regularization if evaluated over a large simulation volume~\cite{kristensen_normalization_2015}.

In terms of these new elastic QNMs, the Green function 
expansion can now be written as
\begin{equation}
\label{eq:greenAns}
{\mathbf{G}}(\mathbf{r},\mathbf{r'};\Omega) = \sum_{{\rm m}=\pm1,\pm2,\cdots}\frac{\tilde{\mathbf{Q}}_{\rm m}(\mathbf{r}){[}\tilde{\mathbf{Q}}_{\rm m}(\mathbf{r'}){]^{T}}}{\rho(\mathbf{r})2{\Omega}(\tilde{\Omega}_{\rm m} - {\Omega})},
\end{equation}
%
where we now only need to perform a sum over just  a few
dominate modes of interest.
We stress that not only does this theory
give a rigorous definition
of Purcell's formula (as we show below), but it fully
accounts for phase effects and non-Lorentzian features from the complex QNMs.

\subsection{Complex mode volumes and
the elastic Purcell formula in terms of QNMs}

The common approach to obtaining mode volumes in the literature \cite{schmidt_elastic_2018,eichenfield_modeling_2009,kipfstuhl_modeling_2014,djafari-rouhani_phoxonic_2016}, is to define the effective mode volume, in the
present case for acoustic modes,  using a normal mode normalization:
\begin{equation}
\label{eq:Veff}
V_{{\rm eff},{\rm m}}^{\rm NM} = \ddfrac{\int d\mathbf{r}  |{\tilde{ \mathbf{ Q}}_{{\rm m}}}(\mathbf{r})|^2}{\rm{max}[|{\tilde{\mathbf{Q}}_{{\rm m}}}(\mathbf{r})|^2]} ,
\end{equation}
where $\mathbf{\tilde Q}_{\rm m}$ is actually the QNM spatial profile (solution with outgoing boundary conditions), but without the normalization of Eq.~\eqref{norm}. However, Eq~\eqref{eq:Veff} is problematic as the value diverges as a function of space for any  dissipative modes (unless they are lossless). In essence it uses a theory
for the normal modes ${\bf Q}_{\rm m}$, which are solutions to a Hermitian eignevalue,
and so it is incorrect to then use a QNM as the mode solution.

To fix this problem, the QNM normalization condition in Eq.~\eqref{norm} allows 
one to obtain a complex (or generalized) effective mode volume 
for a mechanical mode: 
\begin{equation}
\label{eq:V}
{\tilde{V}_{{\rm eff},{\rm m}}(\mathbf{r})} = \frac{1}{\tilde{\mathbf{ Q}}^2_{{\rm m}}(\mathbf{r})},
\end{equation}
where $\tilde{V}_{{\rm eff,m}} = V_{{\rm eff,m}} + iV^{\rm Im}_{{\rm eff,m}}$,
for a specific mode ${\rm m}$.
Note also that we allow the effective mode volume to be a function of 
space here as it formally characterises the mode strength squared at those positions, and is not directly related to the integrated mode volume (which diverges). {For closed cavities, the field squared happens to be related to the total volume of the mode if evaluated at the field maximum. In QNM theory, the total mode volume is ill defined, as it diverges for any open cavity resonator.} 
Thus one can refer to this quantity as a ``localized mode volume,'' 
where the volume is a useful figure of merit for certain applications of interest.
As we will show later, the phase of the QNM can have profound consequences, and is essential to the general theory.


As explained in the introduction,  this modal Purcell factor theory has been 
 well exploited
in optical cavity physics 
and cavity-QED for decades,  and the
underlying physical insight in terms of cavity mode properties would be hard to underestimate as a design tool. Thus, here we aim to derive the expression for an elastic Purcell factor, similar to the work of
Schmidt {\em et al.}~\cite{schmidt_elastic_2018},
but now in terms of the more appropriate QNM
Green functions.

The {\it mechanical} Purcell factor evaluated of an {\it elastic} emitter (coupled to a QNM), ${\bf f}_{\rm{d}} = {f}_{\rm{d}}\bf{n}$, oriented along direction ${\bf n}$, at some position $\mathbf{r}_0$ and at a frequency $\Omega$ can be written as
\begin{equation}
\label{eq:FpEx}
F_{\rm P}({\bf r}_0,\Omega) = \frac{P({\bf r}_0)}{P_0({\bf r}_0)} =  \frac{\rm{Im}[{\bf f}_{\rm{d}}^\dagger\cdot\mathbf{G}(\mathbf{r}_0,\mathbf{r}_0; \Omega)\cdot{\bf f}_{\rm{d}}]}{\rm{Im}[{\bf f}_{\rm{d}}^\dagger\cdot\mathbf{G}_0(\mathbf{r}_0,\mathbf{r}_0; \Omega)\cdot{\bf f}_{\rm{d}}]},
\end{equation}
where $\mathbf{G}_0$ is the Green function in a homogeneous medium.
{Note that ${\bf G}_0$ includes both transverse and longitudinal modes, whereas the QNM one (${\bf G}$) is for  transverse modes only ($\tilde \Omega \neq 0$), which completely dominates the response for resonant cavity structures.}
One can think of 
Eq.~\eqref{eq:FpEx} as
a generalized enhancement factor or
 generalized Purcell factor, as no mode expansions have
 been performed yet (which is necessary to connect to the usual single mode Purcell formula).
The terms
 $P$ and $P_0$ are the radiated power from a dipole (at position ${\bf r}_0$) in an inhomogeneous and homogeneous medium, respectively.
Using an analogous approach to Poynting's theorem, $P_0$ is 
formally defined from
\cite{Novotny2006,schmidt_elastic_2018}: 
\begin{equation}
\label{eq:P0unit}
P_0({\bf r}_0) =  \frac{dW({\bf r}_0)}{dt} = \frac{\Omega}{2} {\rm Im}[{\bf f}_{\rm d}^* \cdot {\mathbf{u}}( \mathbf{r}_0)],
\end{equation}
which can be derived analytically for an isotropic medium
\cite{schmidt_elastic_2018}:
\begin{equation}
\label{eq:P0}
P_0({\bf r}_0) =  \frac{\Omega^2|{\bf f}_{\rm{d}}|^2\alpha}{12\pi\rho(\mathbf{r}_0)},
\end{equation}
where
\begin{equation}
\label{eq:phi}
\alpha = \frac{1}{2}\varv^{-3}_{\rm l} + \varv^{-3}_{\rm s},
\end{equation}
in which $\varv_{\rm l}$ and $\varv_{\rm s}$ are the scalar longitudinal and shear speed of sound in the material, respectively. We can therefore write Eq.~\eqref{eq:FpEx} as
\begin{equation}
\label{eq:Fp2}
F_{\rm p}({\bf r}_0,\Omega) = \eta_{\bf n}\frac{6\pi\rho(\mathbf{r}_0)\rm{Im}[{\bf f}_{\rm{d}}^\dagger\cdot\mathbf{G}(\mathbf{r}_0,\mathbf{r}_0; \Omega)\cdot{\bf f}_{\rm{d}}]}{\Omega|{f}_{\rm{d}}|^2\alpha}.
\end{equation}
{where we have introduced a numerically determined constant $\eta_{\bf n}$ (which depends on the dipole direction in general) to account for elastic anisotropy of the medium; for relatively isotropic crystals, $\eta_{\bf n} \approx 1$.}

The medium 
Green function can now be used with our generalized effective mode volume $\tilde{V}_{\rm eff}$ and Eq.~\eqref{eq:Fp2} for a multi-mode approximation of a cavity decay rate enhancement, and for various other problems in optomechanics. We thus obtain the 
multi-QNM Green function expansion evaluated at a point $\mathbf{r}_0$ within the resonator:
\begin{multline}
\label{eq:FpGq}
F_{\rm p}({\bf r}_0,\Omega) =  \eta_{\bf n}\frac{6\pi\rho(\mathbf{r}_0)}{\Omega|f_d|^2{\alpha}} \times \\ \rm{Im}\left [{\bf f}_{\rm{d}}^\dagger\cdot\sum_{{\rm m}}\frac{\tilde{\mathbf{Q}}_{\rm m}(\mathbf{r}_0){[}\tilde{\mathbf{Q}}_{\rm m}(\mathbf{r}_0){]^{T}}}{\rho(\mathbf{r}_0)2{\Omega}(\tilde{\Omega}_{\rm m} - {\Omega})}\cdot{\bf f}_{\rm{d}}\right].
\end{multline}
 For a slightly more familiar form, we can write this in terms of the (complex) effective mode volume as:
\begin{multline}
\label{eq:FpG}
F_{\rm p}({\bf r}_0,\Omega) =  \eta_{\bf n}\frac{6\pi\rho(\mathbf{r}_0)}{\Omega{\alpha}} \times \\ \rm{Im}\left [\sum_{{\rm m}}\frac{1}{\rho(\mathbf{r}_0)2{\Omega}(\tilde{\Omega}_{\rm m} - {\Omega})\tilde{V}_{{\rm m},{\rm eff}}(\mathbf{r}_0)}\right],
\end{multline}
where we have assumed that the force direction is along the dominant polarization component of the mode, which is 
the usual assumption in Purcell's formula.

Finally, using Eq.~\eqref{eq:Fp2}, we have the 
enhanced emission rate for a single QNM: 
\begin{equation}
\label{eq:Fp9}
F_{\rm p}({\bf r}_0) = \eta_{\bf n}\frac{6\pi Q_{{\rm m}}}{\Omega^3_{{\rm m}}{\alpha} V_{{\rm eff},{\rm m}}({\bf r}_0)},
\end{equation}
in which we assume that the Green function's response is dominated by a single mode, and the response is on-resonance ($\Omega=\Omega_{\rm m}$).
Equation \eqref{eq:Fp9} is the elastic Purcell factor evaluated at the resonant mode $\Omega_{\rm m}$ at the source point, whereas Eq.~\eqref{eq:FpG} is generalized for various positions and frequencies. 
Our expression is consistent with the expression 
recently presented by \cite{schmidt_elastic_2018}, but with a ``corrected'' and {\em generalized} effective mode volume, consistent for
an open cavity mode with
complex eigenfrequencies
and an unconjugated norm.
Note also that our Green
function explicitly includes
the QNM phase, which we show below
is  essential for describing the
response function of several 
QNMs, which can yield
highly non-Lorentzian lineshapes.
 {For background media that are relatively isotopic, $\eta_{\bf n} \approx 1$ (as we find for crystalline Si below), and
 one can simply drop this factor from 
 the Purcell formula and main equations. }

\section{Numerical Results}
\label{sec:results}
\begin{figure}[hbt]
    \centering
    \includegraphics[width=0.48\textwidth]{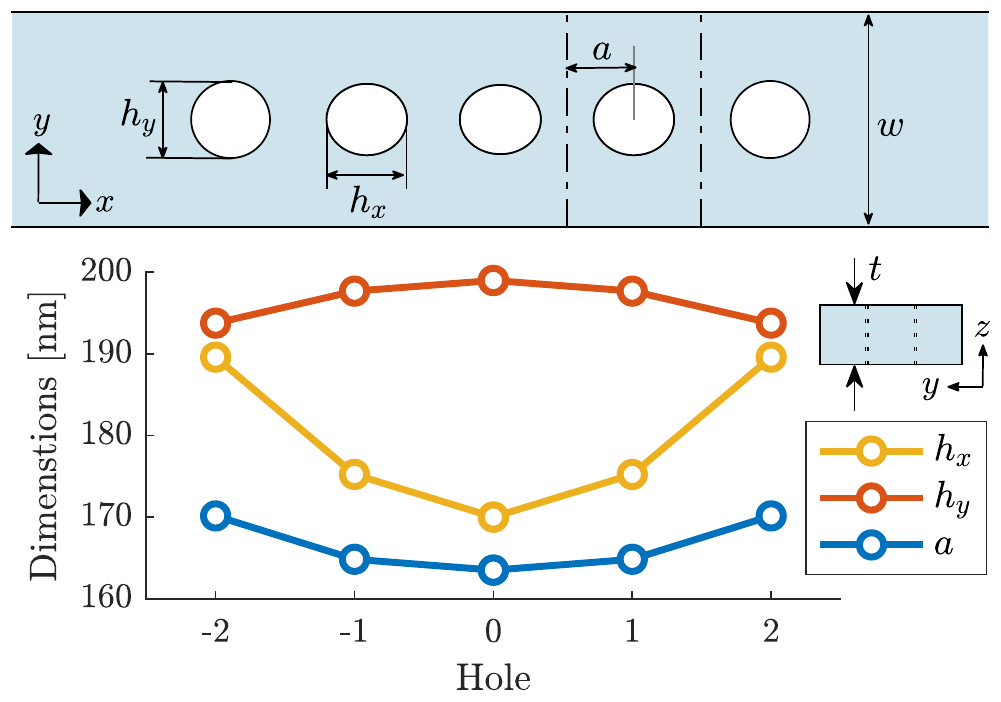}
    \caption{Design specifications of the optomechanical nanobeam cavity, with unit cell parameters $a$, $h_x$, and $h_y$. Dimensions: $w= 0.53\,\rm{\mu}m$ and $t= 0.22\,\rm{\mu}m$. The beam is simulated to be infinitely long using perfectly matched layers. The origin of this coordinate system, $(x,y,z) = (0,0,0)$, is placed at the center of hole-0 at half beam thickness. The 3-hole cavity design uses the holes: -1, 0, and 1. 
    }
    \label{fig:dim}
\end{figure}

\begin{figure*}[t]
    \centering
        \includegraphics[width=1\textwidth]{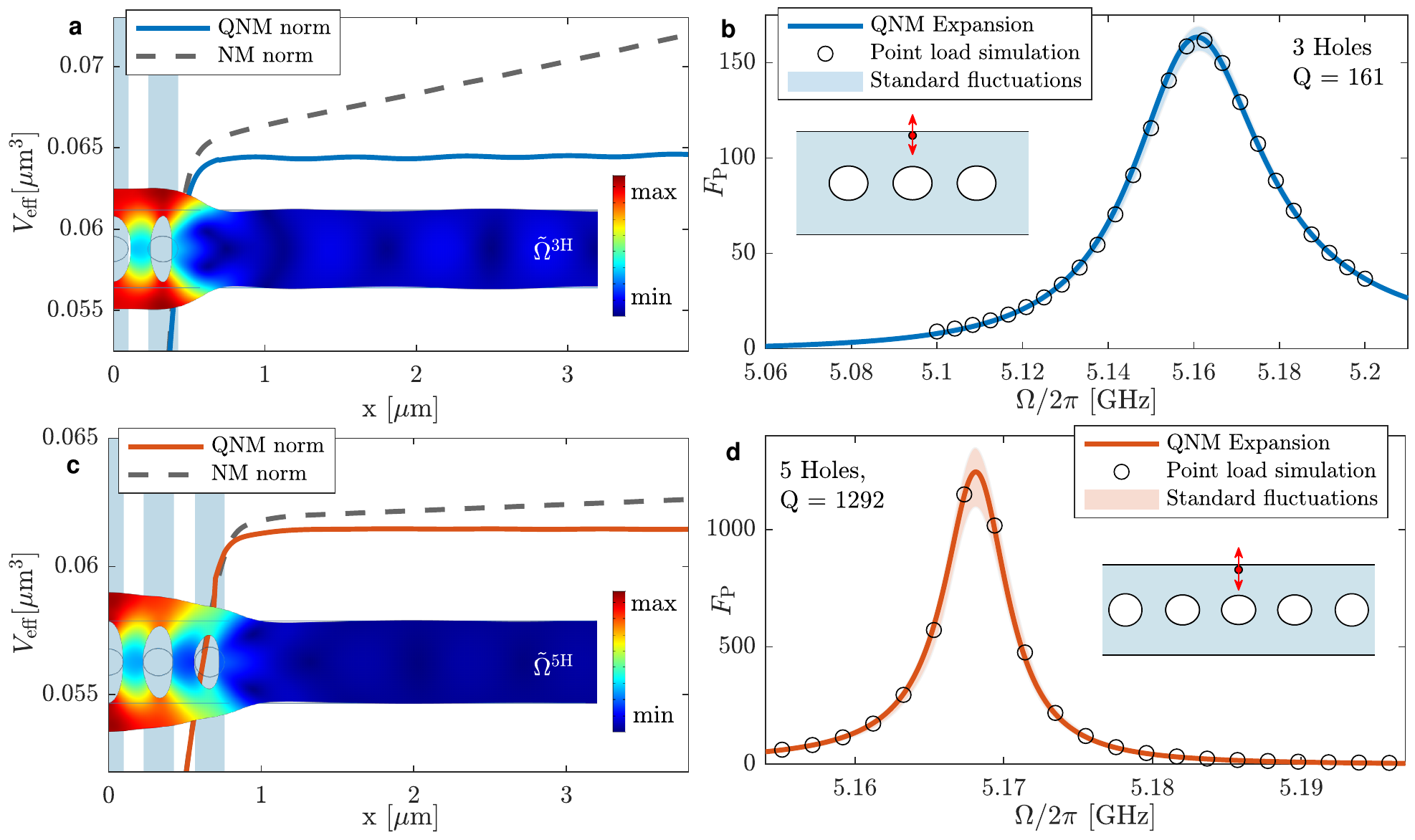}
    \caption{\textbf{a},\textbf{c} Real part of the complex effective mode volume $V_{{\rm eff,m}}$ using the QNM mode normalization from Eq.~\eqref{norm} (solid line) and normal mode (`NM') normalization from Eq.~\eqref{eq:Veff} (dashed) for a 3-hole and 5-hole nanobeam OMC cavity, respectively, {evaluated at ${\bf r}_0=(0.0,235.5,0.0)$ nm)}. Blue shading indicates hole positions with respect to the shown mode profile, which is symmetrical about the $y$-axis at the cavity center. \textbf{b,d} Enhancement rate $F_{\rm P}$ calculated using the Green function expansion (Eq.~\eqref{eq:FpG}) {(with an isotropic material approximation $\eta_{\bf n}=1$)} for the QNMs of interest (solid line). Numerical exact (e.g., with no approximations) Purcell factor Eq.~\eqref{fexact}) obtained from averaged numerical point load simulation from COMSOL (see text). Shaded lines show plus/minus one standard deviation of the numerical solution with mesh sensitivity included.}
    \label{fig:modes}
\end{figure*}

\subsection{Mechanical mode effective mode volumes and Purcell factors}
%
%

%
%


We now corroborate our mechanical QNM theory against rigorous  numerical calculations obtained for fully
three-dimensional mechanical cavities of practical interest in optomechanics.  
{Naturally, the above theory can easily be applied to two dimensional and one dimensional systems as well, with appropriate changes for the background Green functions \cite{kristensen_generalized_2012,kristensen_modeling_2020}.}

In this work, we are interested in modelling optomechanical crystal (OMC) cavities on dielectric nanobeams. We consider the impressive nanobeam structure developed by 
Painter and collaborators~\cite{chan_optimized_2012}, consisting of periodic holes with a lattice taper region, in which the hole spacing and size changes. The taper region causes a Fabry-P\'erot like effect, resulting in spatially overlapping localized optical and mechanical modes. We adapt the original structure to produce higher loss QNMs by using only a small number of holes to form the OMC cavity. This is so that we may test our QNM normalization at low-$Q$, where the normal mode approximation breaks down more dramatically.

 The nanobeam is modeled as anisotropic silicon in free space with elasticity matrix elements $(C_{11},C_{12},C_{44}) = (166,64,80)$ GPa. We consider two OMC cavities, one consisting of 3 holes (3H), and one of 5 holes (5H). Cavity design parameters are specified in Fig.~\ref{fig:dim} along with the beam dimensions. Each cavity design exhibits single mode behaviour over the frequency ranges shown in Fig.~\ref{fig:modes}. We conduct our numerical investigations using the finite element analysis (FEM) commercial software, COMSOL. The beam is simulated to be infinitely long by implementing perfectly matched layers (PMLs), which simulate outgoing boundary conditions. 
 


Employing the eigenfrequency solver, we obtain the dominant QNMs of interest for each cavity at $\Tilde{\Omega}^{3\rm{H}}/2\pi = 5.160 - i0.016$ GHz and $\Tilde{\Omega}^{5\rm{H}}/2\pi = 5.168 - i0.002$ GHz. The spatial profile of each mode is shown in Figs.~\ref{fig:modes}\textbf{a} and \ref{fig:modes}\textbf{c}. We evaluate the effective mode volume of the 3H and 5H QNMs at {${\bf r}_0 = (0.0,235.5,0.0)$ nm (where there is strong localization).} Using Equation~\eqref{eq:V} with the QNM normalization in Eq~\eqref{norm}, the real part of the generalized effective mode volume $\tilde{V}_{\rm eff}$ is plotted as a function of simulation size against the more commonly used normal mode $V_{\rm eff}^{\rm NM}$ (Note that the surface term in Eq~\eqref{norm} is only applied outside of the hole region as it assumes a constant outgoing medium). The normal mode normalization in Eq~\eqref{eq:Veff} assumes that the mode is localized in space. Consequently, when applied to any cavity mode with finite leakage, $V_{\rm eff}^{\rm NM}$ will diverge exponentially when integrated over all space. For less leaky cavities, this divergence is initially quite slow for a small simulation size, whereas the effect is more dramatic for low-$Q$ modes. The QNM normalization, in contrast, converges  as the calculation domain is increased, though may eventually oscillate around the correct value and require regularization~\cite{kristensen_generalized_2012,kristensen_normalization_2015}. Using a sufficient calculation domain size, we obtain {$\tilde{V}_{\rm eff}^{3{\rm H}}({\bf r}_0) = 0.06487 + 0.00943i$ $\mu$m$^3$} and $\tilde{V}_{\rm eff}^{5{\rm H}}({\bf r}_0) = 0.06155 - 0.00355i$ $\mu$m$^3$.

We now make use of the generalized effective mode volume to investigate the elastic Purcell effect. The numerically exact Purcell factor is calculated from 
\begin{equation}
    F_{\rm P}^{\rm exact}({\bf r}_0) = \frac{P_{\rm inhom}({\bf r}_0)}{P_0},
    \label{fexact}
\end{equation}
where $P_{\rm inhom}$ is the power emission from a point load in the inhomogeneous structure (in this case the beam cavity), and $P_0$ from a homogeneous sample. We employ the frequency domain solver in COMSOL to obtain full numerical calculations (plotted in Figure~\ref{fig:modes} \textbf{b,d}). See Appendix~\ref{sec:appendix} for details on the numerical simulations.

Figures~\ref{fig:modes}\,\textbf{b,d} show the semi-analytical enhancement rate $F_{\rm P}(\Omega)$ calculated using Eq.~\eqref{eq:FpG} with $\tilde{V}_{\rm eff}^{3{\rm H}}({\bf r}_0)$ and $\tilde{V}_{\rm eff}^{5{\rm H}}({\bf r}_0)$ for the dominant QNM modes of interest, $\Tilde{\Omega}^{3\rm{H}}$ and $\Tilde{\Omega}^{5\rm{H}}$, respectively. In order to determine $\eta_{\bf n}$, anisotropic material parameters are applied to a numerical model of a homogeneous sample in which the simulation parameters produce a power spectrum agreeing with the analytical expression in Eq.~\eqref{eq:P0} when isotropic material parameters are used. {For our case, we use (and have verified) the isotropic approximation of  $\eta_{\bf n} = P_0^{\rm iso}(\Omega)/P_0^{\rm aniso}(\Omega) \approx 1$, where $P_0^{\rm iso}$ and $P_0^{\rm aniso}$ are the radiated power from a point load in an isotropic and anisotropic homogeneous medium, respectively.} Also plotted in Fig.~\ref{fig:modes}\,\textbf{b,d} is the numerically obtained Purcell factor $F_{\rm P}^{\rm exact}$ calculated from fully three-dimensional frequency domain point load simulations. To account for numerical fluctuations (see Appendix~\ref{sec:appendix}), the average of multiple simulations were conducted in which the mesh parameters are slightly changed. 
Within this fluctuation, we find excellent agreement between the full numerical simulations and the semi-analytical Green function expansion, with the single-mode approximation being sufficient in describing the dominant resonance response of the system. It is worth mentioning that the modal description provides the enhancement of an emitter at any location and frequency given a sufficient number of QNMs (and often just one QNM) which can be calculated in minutes on a single computer. In contrast, normally one must perform a frequency domain calculation for one frequency at a single point load position, which can take anywhere from hours to days for a sufficient spectrum.

\subsection{Coupled mechanical quasinormal modes}
\begin{figure*}[th]
    \centering
    \includegraphics[width=1\textwidth]{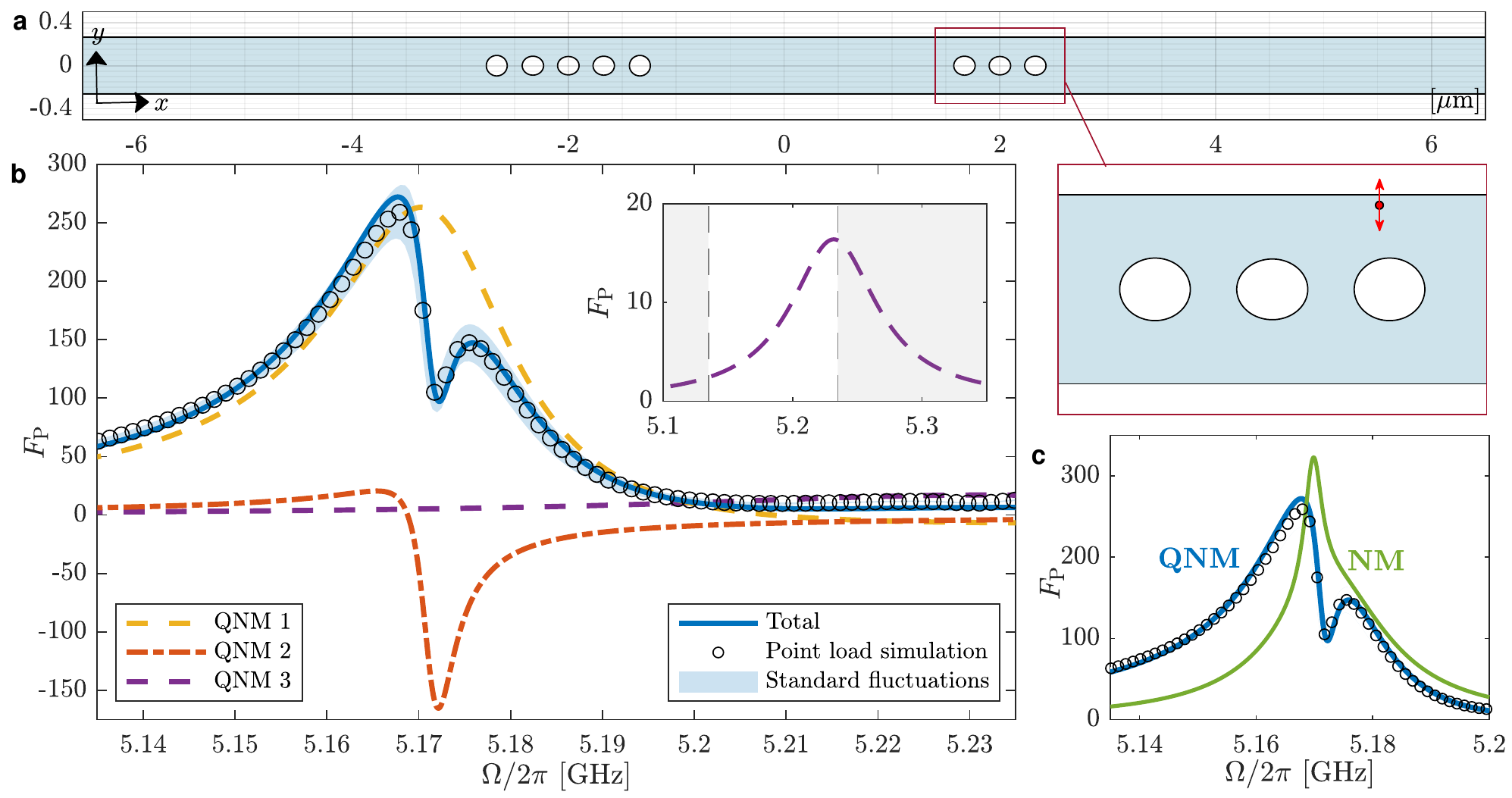}
    \includegraphics[width=1\textwidth]{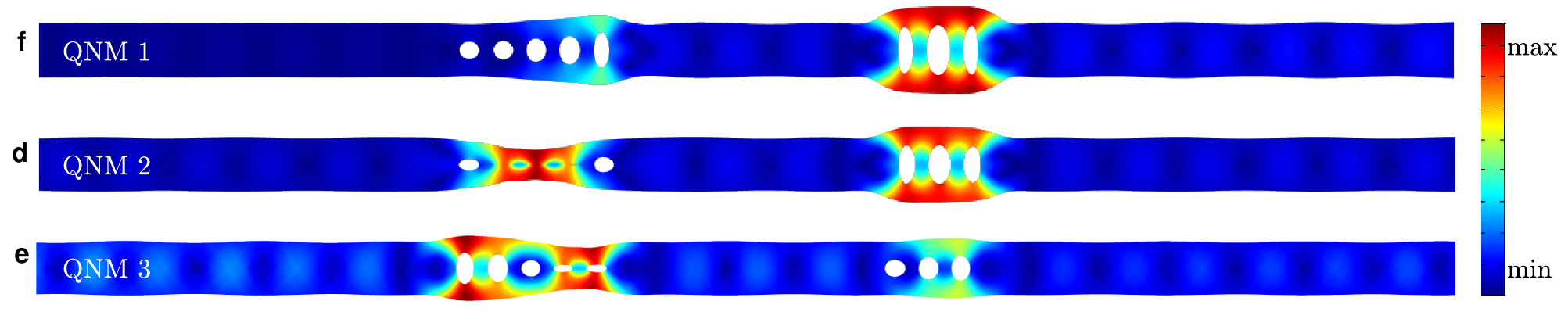}
    \caption{
    \textbf{a} Optomechanical crystal beam geometry, consisting of a 3-hole and a 5-hole cavity seperated by 4 $\mu$m. Zoom-in box shows the simulated point load orientation (aligned with the breathing mode's dominant polarization) and position at ${\bf r}_0 = (2300.0,235.0,0.0)$ nm. \textbf{b} Purcell simulation with a 3-mode approximation of the total decay rate at the same ${\bf r}_0$ (solid line) calculated using Eq.~\ref{eq:FpG} {(using an isotropic material approximation $\eta_{\bf n}=1$)} . Contributions from each individual modes are shown with dashed lines. Inset shows a clearer view of QNM 3, where un-shaded region indicates the frequency range of main plot. Circles show averaged numerical FEM calculations of $F_{\rm P}^{\rm exact}$ (Eq.~\ref{fexact}), with the shaded region showing +/- one standard deviation. \textbf{c} Total decay rate calculated using the normal mode approximation (where the effective mode volume is obtained from Eq.~\ref{eq:Veff}), compared with FEM numerical solution. \textbf{d,e,{f}} Mode profiles of QNM 1 ($\tilde{\Omega}_{\rm 1}/2\pi = 5.172- i0.012$ GHz), QNM 2 ($\tilde{\Omega}_{\rm 2}/2\pi = 5.171- i0.002$), {and QNM 3 ($\tilde{\Omega}_{\rm 3}/2\pi = 5.232 - i0.041$ GHz)},  respectively. Color bar indicates the minimum and maximum displacement relative to each individual mode. Note that these modes are not the original modes, but hybrid modes including the dissipation-induced coupling.
    }
    \label{fig:coupling}
\end{figure*}

Having demonstrated the validity of the generalized effective mode volume with the single mode approximation, we now consider coupled QNMs. 
The optomechanical cavity used above can support both high-$Q$ and low-$Q$ modes, depending on the design and quantity of holes used. We consider an OMC nanobeam structure with two cavities, one designed for relatively high-Q modes (5H), and one for low-Q modes (3H). 

Figure~\ref{fig:coupling}\textbf{a} shows the nanobeam structure, using the same cavity design parameters outlined in Fig.~\ref{fig:dim} and a separation of 4 $\mu$m between the two cavities. We look at two resonant QNMs of interest that are close in frequency with spectral overlap. The first of which, QNM 1 with eigenfrequency $\tilde{\Omega}_{\rm 1}/2\pi = 5.172- i0.012$ GHz ($Q_1 = 216$), is dominated by the 3-hole cavity (mode profile is shown in Figure~\ref{fig:coupling}\textbf{d}). The second, QNM 2 with $\tilde{\Omega}_{\rm 2}/2\pi = 5.171- i0.002$ GHz ($Q_2 = 1293$), is dominated by the 5-hole cavity (Fig.~\ref{fig:coupling}\textbf{e}).

We evaluate the the generalized effective mode volume near the antinode of the 3-hole cavity at ${\bf r}_0 = (2300.0,235.0,0.0)$ nm, and use the multi-QNM Green function expansion to describe the frequency response at this position in Fig.~\ref{fig:coupling}\textbf{b}. Here we seem the hybridization of the individual modes studied earlier ($\Tilde{\Omega}^{3\rm{H}}$ and $\Tilde{\Omega}^{5\rm{H}}$), where QNM 2 exhibits a Fano-resonance that results in an interference effect in the total decay rate. Note that we have used a 3rd mode in our approximation, QNM 3 ($\tilde{\Omega}_{\rm 3}/2\pi = 5.232 - i0.041$ GHz, $\tilde{V}_{\rm eff, 3}({\bf r}_0) = -0.280 - 0.006$ $\mu$m$^3$), to allow for a total decay rate that is positive and well behaved in a large frequency range of interest (while the total decay rate is physically meaningful, contributions from individual modes may not always be); note, however, this mode has negligible contribution to the main dominant response (between 5.15 and 5.19 GHz), and the two-QNM description sufficiently describes the system response. Indeed, we once again have excellent agreement with FEM point load simulations (see Fig.~\ref{fig:coupling}\textbf{b}).

For a more intuitive understanding of the role of the mode phase $\phi_{\rm m} {(\bf r_0) = \rm arg(\tilde{\bf Q_{\rm}}(\bf r_0))}$, with two QNMs,  we can write Eq.~\eqref{eq:FpG} in terms of the two dominant QNMs (QNM 1 and QNM 2) as:
\begin{align}
\label{eq:Fcos+sin}
    &F_{\rm p}({\bf r}_0,\Omega)|_{\rm cos+sin} =
 \frac{3\pi^2\eta_{\bf n}}{\Omega^2\alpha} \times \nonumber \\
 &\left (  \left[\cos 2\phi_1({\bf r}_0)
 + \frac{\Omega_1-\Omega}{\gamma_1}\sin 2\phi_1({\bf r}_0)\right] |{\tilde{\bf Q}_1}({\bf r}_0)|^2\, L_1(\Omega)  \nonumber \right . \\
& + \left .   \left[\cos 2\phi_2({\bf r}_0)
+ \frac{\Omega_2-\Omega}{\gamma_2}\sin 2\phi_2({\bf r}_0)\right] |{\tilde{\bf Q}_2}({\bf r}_0)|^2\,  L_2(\Omega)
 \right ),
\end{align}
where we use the normalized Lorentzian function,
\begin{equation}
     L_{\rm m}(\Omega)
     = \frac{\gamma_{\rm m}/\pi}{(\Omega_{\rm m}-\Omega)^2+\gamma_{\rm m}^2},
\end{equation}
and we assume the force dipole is projected along
the dominant field direction, namely
$|\tilde{\bf Q}|^2 = |{\bf \tilde Q \cdot {\bf n}}|^2$, 
though this can easily be generalized.
To better clarify the underlying physics of the various
phase terms, we also define two other functions,
one that neglects the $\sin$
contributions:
\begin{align}
\label{eq:Fcos}
    F_{\rm p}({\bf r}_0,\Omega)|_{\rm \cos} &=
 \frac{3\pi^2\eta_{\bf n}}{\Omega^2\alpha} \times \nonumber \\
 &\big( \cos 2\phi_1({\bf r}_0)
 |{\tilde{\bf Q}_1}({\bf r}_0)|^2\,L_1(\Omega) \nonumber \\
+  &\cos 2\phi_2({\bf r}_0)
 |{\tilde{\bf Q}_2}({\bf r}_0)|^2\,L_2(\Omega)
 \big), 
\end{align}
and one that neglects the phase completely:
\begin{align}
\label{eq:Fabs}
    &F_{\rm p}({\bf r}_0,\Omega)|_{\rm abs} =
    \nonumber \\ & \ \ \ \frac{3\pi^2\eta_{\bf n}}{\Omega^2\alpha} 
  \left (
 |{\tilde{\bf Q}_1}({\bf r}_0)|^2\,L_1(\Omega)
+ 
 |{\tilde{\bf Q}_2}({\bf r}_0)|^2\,L_2(\Omega)\right ).
\end{align}
All three  agree (Eqs.~\eqref{eq:Fcos+sin},\eqref{eq:Fcos},
\eqref{eq:Fabs}) {\em only} when $\cos(2 \phi_{1})=\cos(2 \phi_{2})=1$,
and $\sin(2 \phi_{1})=\sin(2 \phi_{2})=0$. Note also that Eq.~\eqref{eq:Fabs} has the same form as a normal mode solution, though its effective mode volume is different, and the later will in general be overstated (yielding a smaller
Purcell factor value).

\begin{figure}[hbt]
    \centering
    \includegraphics[width=0.48\textwidth]{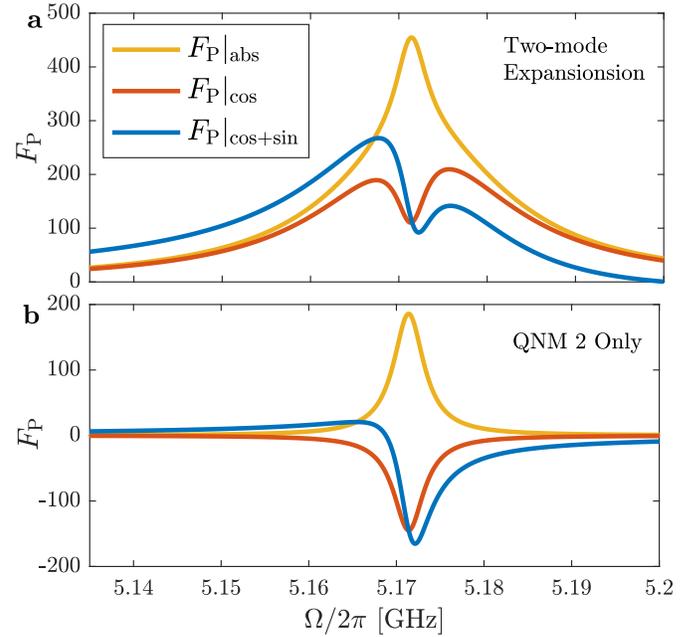}
    \caption{ Two QNM expansion (\textbf{a}) and single mode expansion of QNM 2 (\textbf{b}) using all phase terms (Eq.~\eqref{eq:Fcos+sin}), the cos term only (Eq.~\eqref{eq:Fcos}, and without any phase dependence (Eq.~\eqref{eq:Fabs}).}
    \label{fig:phase}
\end{figure}

For our numerical example,
the QNM phase values at the point of interest are $2\phi_1(\mathbf{r}_0) = 0.3204$ and $2\phi_2(\mathbf{r}_0) = 2.4643$ for QNM 1 and QNM 2, respectively. While the QNM 1 phase shift $\phi_{1}$ is relatively small, the near $180^o$ phase shift of ${\bf \tilde{Q}_2}^2({\bf r}_0)$ results in the negative contribution to the overall enhancement.
From these phase values, the cosine values are $\cos(2 \phi_{1})\approx 0.95$
and $\cos(2 \phi_{1})\approx -0.7$, so the latter will contribute
as a negative Lorentzian lineshape.
Figure \ref{fig:dim} shows the 
Purcell factor predictions from Eqs.~\eqref{eq:Fcos+sin}, \eqref{eq:Fcos},
and \eqref{eq:Fabs}, which clearly demonstrates the role of the phase terms. We can see that neglecting the phase of the mode fails entirely in describing the interference effect (Figure~\ref{fig:phase}\textbf{a}). Considering only the cosine terms still provides a negative contribution from QNM 2 (this is equivalent to only using the real part of $\tilde{V}_{\rm eff, 1}$), however, the asymmetry of the lineshape requires the full phase. In fact, the sin terms are
$\sin(2 \phi_{1})\approx 0.31$
and $\sin(2 \phi_{1})\approx 0.36$, which are significant and certainly cannot be ignored in general. Indeed, the pronounced Fano feature can only be correctly described when the complete phase is used (see Fig.~\ref{fig:phase}\textbf{b}).

Equation~\eqref{eq:Fcos+sin} provides a clear understanding of phase interactions between coupled modes. However, rather than work with the complex phases, it can be convenient to work in the complex effective mode volume picture, i.e Eq.~\eqref{eq:FpG}, which is in the spirit of Purcell's formula. A simple way to do this is to write Eq.~\eqref{eq:Fcos+sin} in terms of the real and imaginary parts of the generalized effective mode volume:
\begin{align}
   & F_{\rm p}({\bf r}_0,\Omega) \approx 
 \frac{3\pi^2\eta_{\bf n}}{\Omega^2\alpha} \times \nonumber \\
 &\left (  \left[ {\rm Re} \left(\frac{1}{\tilde V_{\rm eff,1} ({\bf r}_0)} \right)
 + \frac{\Omega_1-\Omega}{\gamma_1} {\rm Im} 
 \left(\frac{1}{\tilde V_{\rm eff,1}({\bf r}_0)}\right) \right]  L_1(\Omega)  \nonumber \right . \\
& + \left .  \left[ {\rm Re} \left( \frac{1}{\tilde V_{\rm eff,2} ({\bf r}_0)} \right)
 + \frac{\Omega_2-\Omega}{\gamma_1} 
 {\rm Im} \left (\frac{1}{\tilde V_{\rm eff,2}({\bf r}_0)} \right) \right]  L_2(\Omega)
 \right ),
 \label{eq:Fv}
\end{align}
where we have effectively replaced 
the sin and cos terms, as well as $|\tilde{\bf Q}_{\rm m}|^2$, with the real and imaginary parts of $1/\tilde{V}_{{\rm eff},{\rm m}}$.  It is now easier to see that the normal mode solution using the entirely real $V_{\rm eff,m}^{\rm NM}$~\cite{schmidt_elastic_2018} will always be a simple sum of two Lorentzians:
\begin{align}
    F_{\rm p}^{\rm NM}({\bf r}_0,\Omega) &\approx 
 \frac{3\pi^2\eta_{\bf n}}{\Omega^2\alpha} 
 \left (   \frac{ L_1(\Omega) }{V_{\rm eff,1}^{\rm NM} ({\bf r}_0)}
 +  \frac{ L_2(\Omega)}{V_{\rm eff,2}^{\rm NM} ({\bf r}_0)}
 \right ),
 \label{eq:Fnm}
\end{align}
as plotted in Fig.~\ref{fig:coupling}\textbf{c},
which shows a drastic failure 
of the normal mode theory. Note  that Eq.~\eqref{eq:Fnm} for  a  single mode is equivalent to the one in Ref.~\cite{schmidt_elastic_2018}.


The calculated generalized effective mode volumes of the QNMs of are interest are $\tilde{V}_{\rm eff, 1}({\bf r}_0) = 0.058 - 0.019i$ $\mu$m$^3$ and $\tilde{V}_{\rm eff, 2}({\bf r}_0) = -0.399 - 0.321i$ $\mu$m$^3$ for QNM 1 and QNM 2, respectively. The negative nature of $\tilde{V}_{\rm eff}$ at
${\bf r}_0$ is simply a result of the phase shift caused by the interaction between the two cavity resonances. This is also seen with QNMs in optics~\cite{lasson_semianalytical_2015,kamandar_dezfouli_modal_2017}, where the QNM
phase causes the Fano-like resonance.
Using quantized QNMs in quantum optics
gives essentially the same result as classical QNM theory in the bad cavity limit (within numerical precision), but where the interpretation is 
now through off-diagonal mode coupling~\cite{franke_quantization_2019};  here the dissipation-induced interference cannot be explained by normal mode quantum theories such as the dissipative Jaynes-Cumming model. In this regard, it would be very interesting to develop a quantized QNM theory for 
mechanical modes as well.





\section{Conclusions}
We have introduced a QNM formalism  for mechanical cavity modes and shown that the commonly used normal mode description is problematic for cavity resonances with finite loss. Instead, we have presented and employed a complex, position dependant effective mode volume for mechanical modes using a QNM normalization which can be used to solve a  range of force-displacement problems. For validation of the theory, we carried out an analytical Green function expansion using QNMs with the an elastic Purcell factor expression, and found excellent agreement with rigorous numerical simulations for 3D
optomechanical beams. 

We then demonstrated the accuracy of the QNM theory in explaining interference effects of coupled cavity modes, and pointed out the drastic failure of the usual normal mode theory. 
Specifically, we explicitly showed the role of the QNM phase in yielding a
Fano-like resonance and explained this analytically and numerically from 
interference effects that are completely absent in a normal mode theory.
This QNM approach should serve as a robust and valuable tool in the 
understanding and and development of emerging optomechanical technologies.

\label{sec:conclusions}

\acknowledgements

This work was funded by the Natural Sciences and Engineering
Research Council of Canada, the Canadian Foundation for Innovation and Queen's University, Canada.
We gratefully acknowledge Mohsen Kamandar Dezfouli and Chelsea Carlson for useful discussions and help with the COMSOL calculations. This research was enabled in part by computational
support provided by the Centre for Advanced Computing
(\url{http://cac.queensu.ca}) and Compute Canada (\url{www.computecanada.ca}).

\appendix

{
\section{Analysis of quasinormal mode 3}
The third mode included in the total decay rate in Fig.~\ref{fig:coupling}, `QNM 3', has no qualitative influence on the main Fano feature we are modeling. In essence, it merely produces a small background bump in the total decay rate far in frequency from the hybridized modes of interest. However,  without its inclusion we have nonphysical negative enhancement, so it is needed in general to quantitativley connect to the total Purcell factor over a relatively broad frequency range. From our calculations,  QNM 3 is likely a modification of a (second) mode that arises in the single cavity 5-hole structure at a slightly higher frequency than the primary mode of interest, $\tilde{\Omega}^{\rm 5H}$. 

Figure~\ref{fig:secondary} shows (a) the two-mode approximation of the total decay rate of the {\bf single} 5-hole cavity nanobeam at ${\bf r}_0^{\rm 5H}$, along with (b) the spatial profile of QNM 3. This is the same structure used in Fig.~\ref{fig:modes}\textbf{c, d}, in which we used a one-mode approximation using only the primary mode of interest $\Tilde{\Omega}^{5\rm{H}}/2\pi = 5.168 - i0.002$ GHz. Here we also include the nearest resonant mode at $\Tilde{\Omega}_{\rm s}^{5\rm{H}}/2\pi = 5.210 - i0.051$ GHz. With a quality factor of 53 and $\tilde{V}_{\rm eff}^{5{\rm H}, {\rm s}} = 0.229 - 0.124i$ $\mu$m$^3$, its contribution to the total day rate over this frequency range is overshadowed by $\tilde{\Omega}^{\rm 5H}$, showing that  the single-mode expansion in Fig.~\ref{fig:modes}\textbf{d} is a sufficient approximation. In addition to having comparable quality factors and relative distances in frequency from $\tilde{\Omega}^{\rm 5H}$ and QNM 2, respectively, $\Tilde{\Omega}_{\rm s}^{5\rm{H}}$ and QNM 3 have similar mode profiles at the 5-hole cavity region. Leading us to conclude that QNM 3 is simply $\Tilde{\Omega}^{5\rm{H}}/2\pi = 5.168 - i0.002$ GHz adapted to the addition of the 3-hole cavity on the same beam.
\begin{figure}[htb]
    \centering
    \includegraphics[width=0.99\columnwidth]{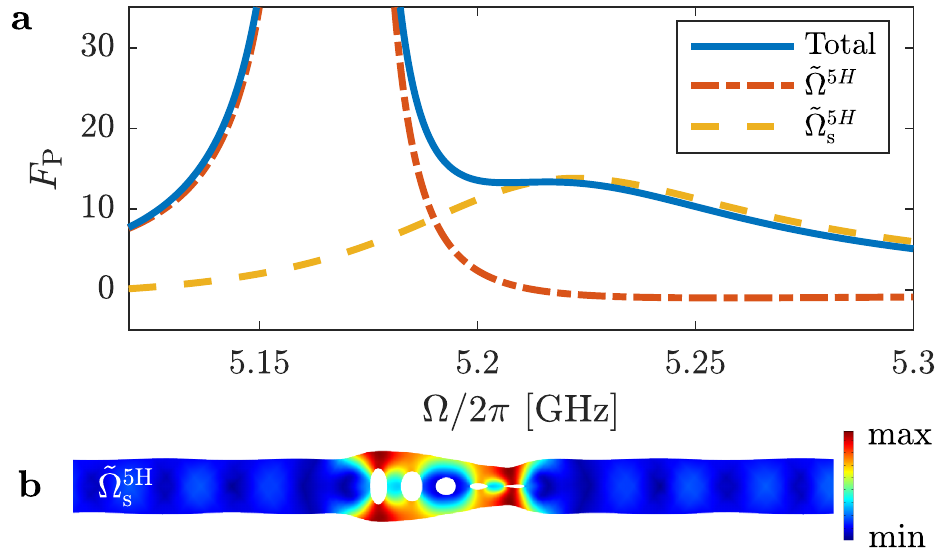}
    \caption{{\textbf{a} Two-mode approximation of the total decay rate of the {\bf single} 5-hole cavity nanobeam at ${\bf r}_0 = (0.0,235.5,0.0)$ nm calculated using Eq.~\eqref{eq:FpG}. Dashed lines show the individual mode contributions of the the primary mode $\Tilde{\Omega}^{5\rm{H}}$ and secondary mode $\Tilde{\Omega}_{\rm s}^{5\rm{H}}$. \textbf{b} Mode profile near the 5 hole cavity of $\Tilde{\Omega}_{\rm s}^{5\rm{H}}$ (QNM 3 in the main text). }
    }
    \label{fig:secondary}
\end{figure}
}

{
\section{Purcell Factor at different positions}
The choice of the evaluation point ${\bf r}_0 = (0.0,235.5,0.0)$ nm is significant since the QNMs of interest for the single cavity structures are relatively strong at this position. Figure~\ref{fig:cross} shows the peak Purcell factor of $\Tilde{\Omega}^{3\rm{H}}$ and $\Tilde{\Omega}^{5\rm{H}}$ modes on two cross sectional lines around this point. The evaluation point for the coupled cavity structure in Figure~\ref{fig:coupling} was chosen simply because it exhibited a pronounced Fano feature. However, for completeness, we also include the total decay rate of the coupled cavity structure evaluated near the 5-hole cavity region in Fig.~\ref{fig:third}.
\begin{figure}
    \centering
    \includegraphics[width=0.99\columnwidth]{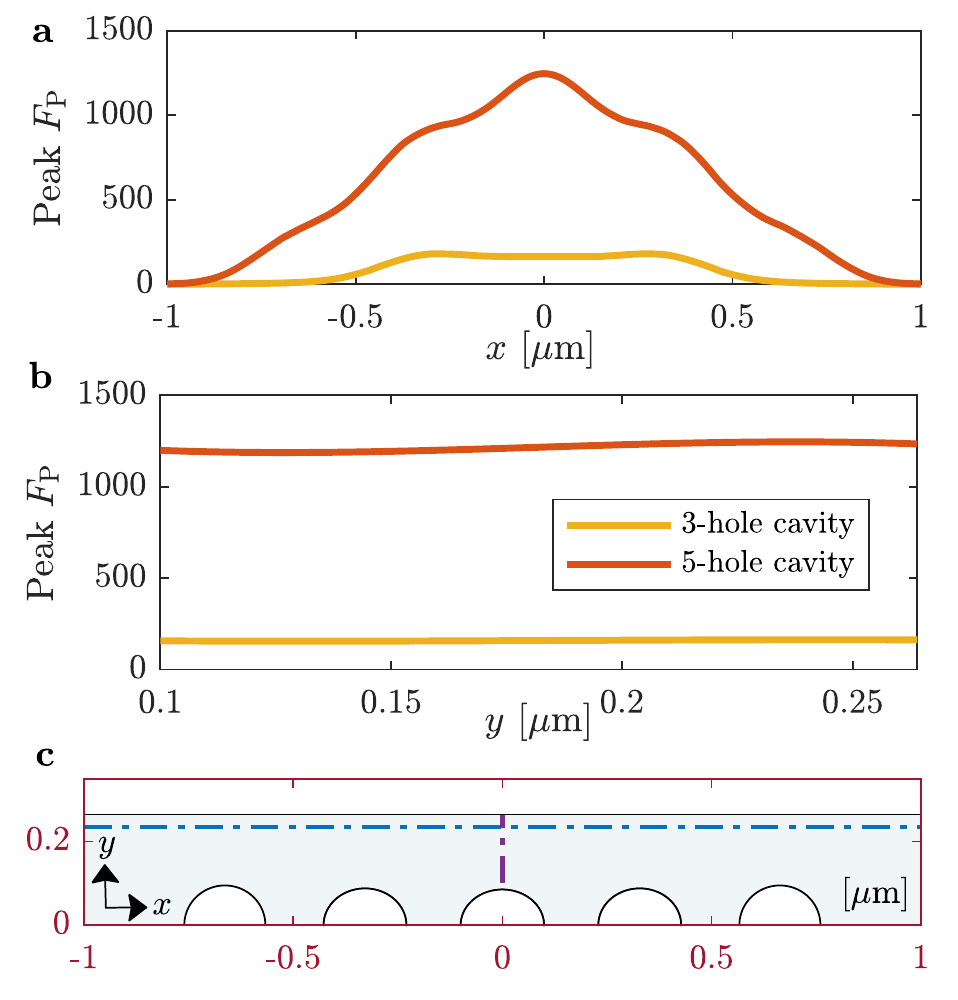}
    \caption{{Peak decay rate enhancement of the dominant QNM for the {\textbf{a}} 3-hole cavity ($\Tilde{\Omega}^{3\rm{H}}$) and {\textbf{b}} 5-hole cavity ($\Tilde{\Omega}^{5\rm{H}}$) at a range of $x$ and $y$ evaluation points in the center of the beam ($z = 0$ $\mu$m). {\textbf{c}} shows cross sections at $y = 0.2355$ $\mu$m ({\textbf{a}}, blue dashed) and $x = 0$ $\mu$m ({\textbf{b}}, purple dashed).}}
    \label{fig:cross}
\end{figure}
\begin{figure}[h]
    \centering
    \includegraphics[width=0.99\columnwidth]{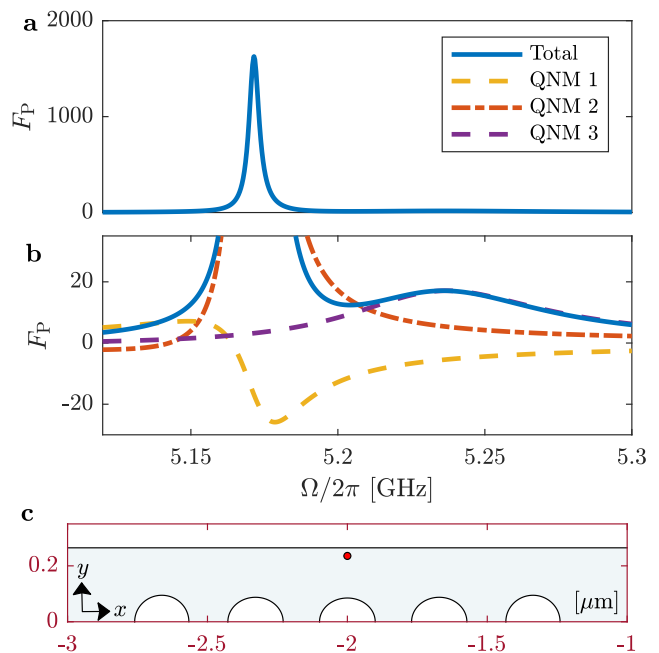}
    \caption{{\textbf{a} 3-mode approximation of the total decay rate near the 5-hole cavity region of the coupled cavity structure (${\bf r}_0 = (-2000.0,235.5,0.0)$ nm). \textbf{b} Zoom-in of {\bf a}, with contributions from each individual modes shown (dashed lines). {\textbf{c}} zoom-in of the beam geometry shown in Figure~\ref{fig:coupling}{\bf a}. Red point indicates the evaluation point at this ${\bf r}_0$.  Note that we see QNM 1 exhibit a negative contribution to the total decay at this position, but it is overwhelmed by the relatively large enhancement of QNM 2.}}
    \label{fig:third}
\end{figure}
}

\begin{figure*}
    \centering
    \includegraphics[width=0.70\textwidth]{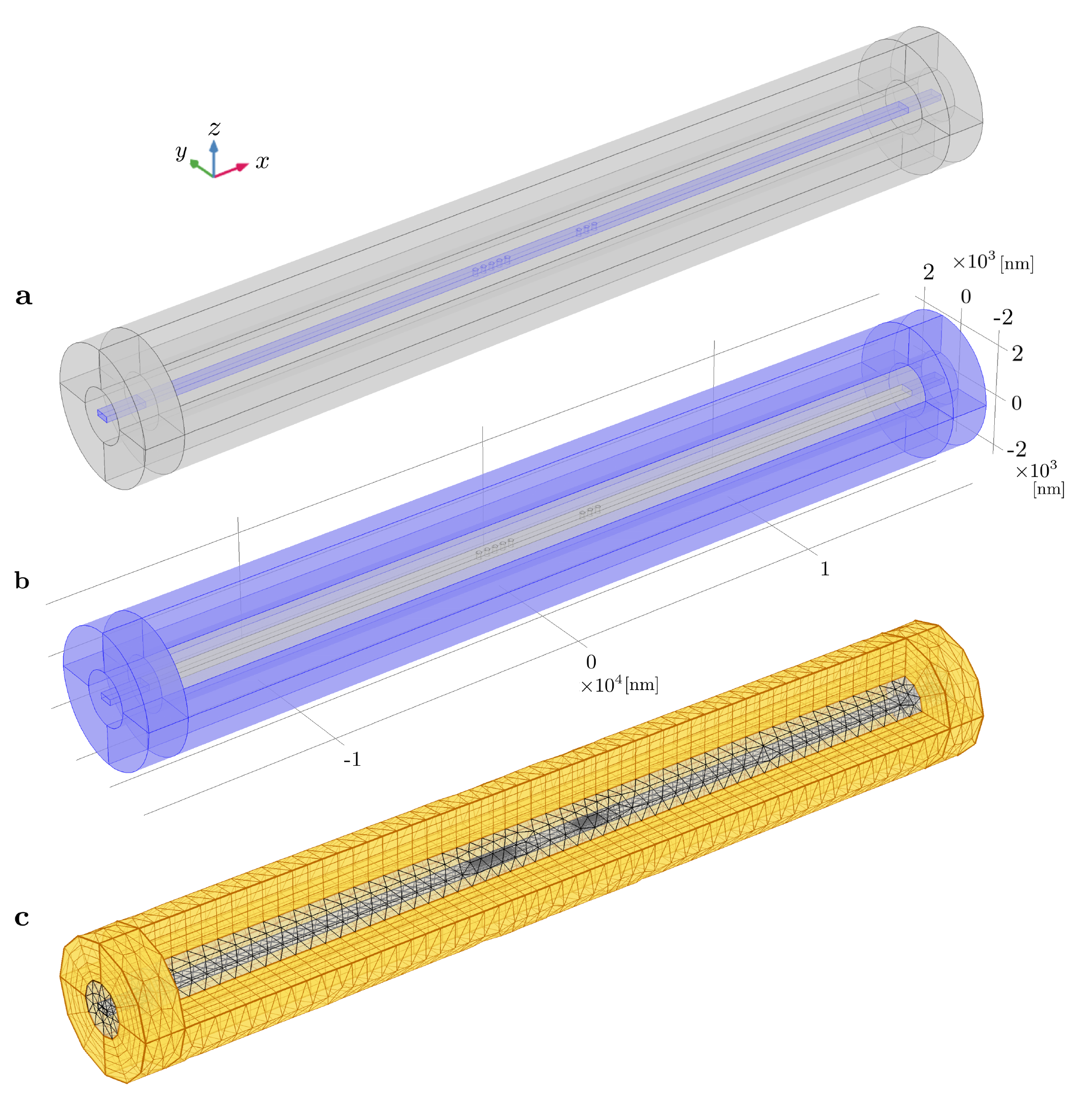}
    \caption{ The COMSOL simulation geometry used for coupled modes consists of 30.5 $\mu$m long nanobeam (see Figs.~\ref{fig:dim} and \ref{fig:coupling}\textbf{a} for beam design) with a 2.25 $\mu$m PML in the $x$ direction. The radial PML is 1.5 $\mu$m thick and is separated by a 1 $\mu$m buffer from the beam center. \textbf{a} Domain highlighted in blue indicates the embedded nanobeam with silicon material parameters. The surrounding domains (grey) use fictitious material parameters approximating a vacuum for inhomogeneous simulations, and use silicon material parameters for the homogeneous simulation. \textbf{b} Regions highlighted in blue show the PML domain, where we have cut out a quadrant of the radial PML to show the interior domains. \textbf{c} Domains highlighted in yellow show the radial PML regions, using a swept mesh with 5 layers. Remaining domains (grey) use a free tetrahedral mesh (see Table~\ref{tab:meshP} for mesh parameters).
    }
    \label{fig:pmlSelection}
\end{figure*}

\section{COMSOL calculations and numerical  Purcell factors}
\label{sec:appendix}


The power emission spectrum is obtained numerically by integrating the mechanical flux $\bm{I}$ over a small sphere around the point load. The mechanical flux is given by:
\begin{equation}
    \bm{I} = -\bm{\sigma}\cdot\bm{v},
    \label{Pflux}
\end{equation}
where $\bm{v}$ is the velocity vector. Power flow calculations in
COMSOL were found to be 
sensitive to mesh geometry, with $P_{\rm inhomo}$ (nanobeam cavity) and $P_0$ (homogeneous bulk material)
fluctuating dramatically
with small changes in mesh parameters. However, $P_{\rm inhom}/P_0$ was found to be
more convergent provided that the mesh geometry in the simulation of the beam be exactly identical to the simulation mesh of the homogeneous medium. This was achieved by using the same geometry and mesh points for both $P_{\rm inhom}$ and $P_0$ simulations, with the material parameters surrounding the beam (see Figure~\ref{fig:pmlSelection}\textbf{a}) changed from silicon (homogeneous case) to a fictitious material with elasticity $(C_{11},C_{12},C_{44}) = (0,0,0)$ GPa and a density of $0.001$ ${\rm kg}/{\rm m}^3$ in order to approximate a vacuum (inhomogeneous case). The fictitious material is necessary as meshes in the COMSOL solid mechanics solver must be assigned a material, and the chosen elasticity and density for the vacuum approximation suffice. In fact, we found that using any density less than $ 0.01$ ${\rm kg}/{\rm m}^3$ has negligible effect on the calculated $P_{\rm inhom}$ and the calculated eigenfrequencies of the QNMs (which agree with the in-vacuum simulations). 
\begin{table}[h]
    \centering
    \begin{tabularx}{0.99\columnwidth} { 
  | >{\raggedright\arraybackslash}X 
  | >{\centering\arraybackslash}X 
  |}
    \hline
    Max. element size & 1000 [nm]\\ 
    \hline
    Min. element size & 0.1 [nm]\\
    \hline
    Max. element growth rate & 1.5 \\
    \hline
    Curvature factor & 0.6 \\
    \hline
    Resolution of narrow regions & 0.5 \\
    \hline
    Max. point load element size & 0.2 [nm]\\

    \hline
    \end{tabularx}
    \caption{COMSOL simulation mesh parameters used.}
    \label{tab:meshP}
\end{table}
\begin{figure}[h]
    \centering
    \includegraphics[width=0.99\columnwidth]{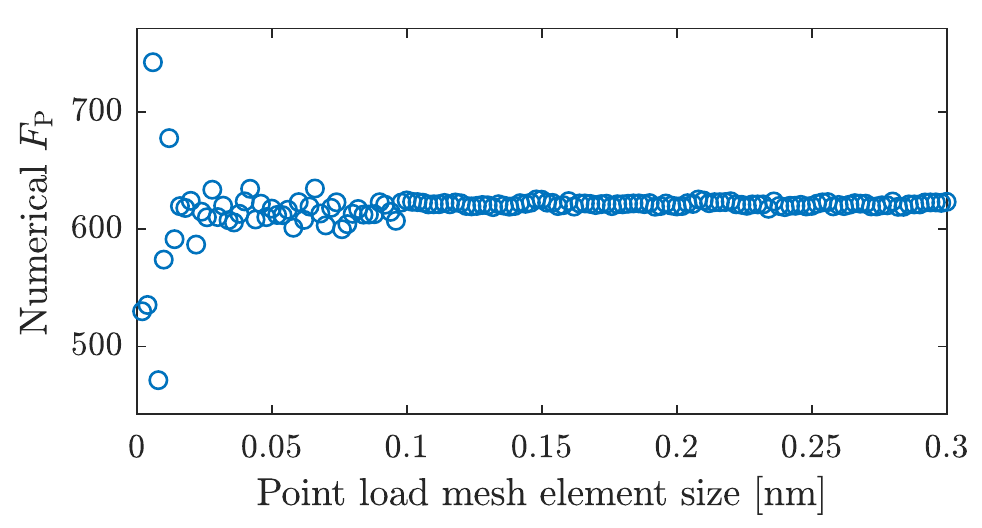}
    \caption{COMSOL simulations of the numerical Purcell factor (Eq.~\ref{fexact}) of a mechanical mode at a single arbitrary frequency and dipole position for various maximum point load mesh element sizes.}
    \label{fig:fluc}
\end{figure}
Table~\ref{tab:meshP} outlines the COMSOL mesh parameters used in our simulations, which were found to give consistent (and computationally feasible) solutions provided that an appropriate mesh element size is used for the point load. For absorbing boundary conditions, PMLs with polynomial coordinate stretching are used with the scaling factor and curvature parameter set to 1.

Figures~\ref{fig:pmlSelection}\textbf{b} and ~\ref{fig:pmlSelection}\textbf{c} outline the PML simulation domains and their meshing type (see figure caption text). Figure~\ref{fig:fluc} shows a parameter sweep of the maximum mesh element size assigned to the point load, where we can see convergence with small fluctuation for element sizes larger than 0.1 nm. For calculations in this work, a point load mesh element size of around 0.2 nm is used. Small shifts in this parameter (in the region of convergence) effectively minutely changes the simulation mesh, resulting in the fluctuations around the solution. In order to account for this, we take the average solution of simulations with slightly varied mesh sizes.   

\clearpage
\newpage
\bibliography{paperRefs}

\end{document}